\documentclass[
twocolumn,showpacs,amsmath,amssymb,aps,prd]{revtex4}
\usepackage{dcolumn}
\usepackage{amsthm,amscd}
\usepackage{amsfonts,bm,latexsym}

\def\be{\begin{equation}}
\def\ee{\end{equation}}
\def\bea{\begin{eqnarray}}
\def\eea{\end{eqnarray}}

\def\nn{\nonumber}
\def\p{\partial}

\begin{document}

\title{General Rotating Charged Kaluza-Klein AdS Black Holes in Higher Dimensions}

\author{Shuang-Qing Wu} \email{sqwu@phy.ccnu.edu.cn}
\affiliation{Institute of Theoretical Physics, China West Normal University, Nanchong,
 Sichuan 637002, People's Republic of China}
\affiliation{College of Physical Science and Technology, Hua-Zhong Normal University,
 Wuhan, Hubei 430079, People's Republic of China}

\begin{abstract}
I construct exact solutions for general nonextremal rotating, charged Kaluza-Klein black holes
with a cosmological constant and with arbitrary angular momenta in all higher dimensions. I then
investigate their thermodynamics and find their generalizations with the NUT charges. The metrics
are given in both Boyer-Lindquist coordinates and a form very similar to the famous Kerr-Schild
ansatz, which highlights its potential application to include multiple electric charges into
solutions yet to be found in gauged supergravity. It is also observed that the metric ansatz
in $D = 4$ dimensions is similar to those previously suggested by Yilmaz and later by Bekenstein.
\end{abstract}


\maketitle

\section{Introduction}
It is generally accepted that an appropriate ansatz for the metric and the gauge potential plays a
crucial role in finding an exact solution to Einstein and Einstein-Maxwell field equations. A well-known
example is provided by the Kerr solution \cite{Kerr}, which was first derived via the Kerr-Schild
ansatz \cite{KS65}. This ansatz was then used by Myers and Perry \cite{MP86} in 1986 to successfully
obtain higher-dimensional vacuum generalizations of the Kerr solution. Several years ago, the Kerr-Schild
form was adopted again by Gibbons, \textit{et al} \cite{GLPP04,GLPP05} to include a cosmological
constant in the vacuum Myers-Perry's solution in all higher dimensions.

With the discovery of the remarkable anti-de Sitter/conformal field theory (AdS/CFT) correspondence,
it is of considerable interest to generalize the above-mentioned neutral rotating solutions to charged
ones in arbitrary dimensions. One of the major reasons is that rotating charged black holes with a
cosmological constant in higher dimensions can provide new important gravitational backgrounds for
the study of the microscopic entropy of black holes and for testing the AdS/CFT correspondence within
the string theory framework. In the case of ungauged supergravity, research on string dualities
has revealed that some global symmetries can be used as solution-generating transformations to obtain
new solutions from old ones. Therefore, it is straightforward to employ a solution-generating
procedure to generate charged solutions from neutral ones. In general, the generated solutions are
very complicated and typically characterized by multiple electromagnetic charges, in addition to
the mass and angular momenta. (For an earlier review, see \cite{Youm99} and references therein.)

However, the situation is quite different for the gauged cases. There is no longer a solution-generating
technique available for deriving the nonextremal charged black holes from neutral solutions in the gauged
supergravity, since the presence of a cosmological constant breaks down the corresponding global symmetries
of ungauged supergravity. One has little option but to resort to brute-force calculations \cite{bfm1,bfm2},
starting from a guessed ansatz to verify that all the equations of motion are completely satisfied. At
present, almost all the previously-known solutions of rotating charged AdS black holes in higher dimensions
have been obtained in this way. What is more, they are limited to very special cases either with some charges
equal, or with equal rotation parameters \cite{CCLP07}.

As far as the simplest case with only one charge is concerned, the currently-known charged nonextremal
rotating black hole solutions within the Kaluza-Klein supergravity theory are as follows. The first
rotating charged black hole in the four-dimensional Kaluza-Klein theory was derived in \cite{FZB87} via
the boost-reduction procedure, and its extension with a NUT charge was obtained recently in \cite{ACD08}.
Generalizations to all higher dimensions were presented in \cite{KMNV06}. In the case of Kaluza-Klein
gauged supergravity, a single-charged solution with only one rotation parameter non-vanishing in five
dimensions was found for the first time in \cite{CCLP05}, and the general rotating charged solution with
only one charge nonzero and with two unequal rotation parameters was then announced in \cite{CCLP07}.
Inspired by the work \cite{CCLP05}, Chow \cite{Chow11} recently found a solution describing a rotating
charged AdS(-NUT) black hole in four-dimensional Kaluza-Klein gauged supergravity. General solutions
that describe rotating charged Kaluza-Klein-AdS (KK-AdS) black holes in $D \geq 6$ dimensions are not
yet known explicitly until the present work. For the most general case with three unequal charges and with
two independent rotation parameters, the explicit form of charged rotating AdS$_5$ solutions has remained
unknown up to now.

So far, all the previously-obtained solutions for rotating charged AdS black holes were not derived via
a universal method other than via a combination of guesswork and trial and error, followed by explicit
verification of the field equations. A natural question is, can one develop an effective method
somewhat like the Kerr-Schild ansatz to overcome the difficulty in the construction of rotating black
holes with multiple charges in gauged supergravity? The answer seems likely to be definitive. The purpose
of this article is to present a clue to resolve this dilemma. As a first step towards this direction and
for simplicity, I shall be mainly concerned with the single-charge case in Kaluza-Klein supergravity.

In this paper, I begin by presenting the general solutions for rotating, charged KK-AdS black holes
with single electric charge and with arbitrary angular momenta in all higher dimensions. Then I calculate
the conserved charges that obey the first law of thermodynamics, and make a generalization to include
the NUT charges. After these, a primary analysis of the metric structure is given, which sheds new
light on constructing the most general rotating, multiple-charged AdS black hole solutions yet unknown
in gauged supergravity theories.

\section{General KK-AdS solution}
To present the general exact solutions, conventions are adopted as those in \cite{GLPP04}. Let the
dimension of spacetime be $D = 2N +1 +\epsilon \geq 4$, with $N = [(D-1)/2]$ being the number of
rotation parameters $a_i$, and $2\epsilon = 1 +(-1)^D$. Let $\phi_i$ be the $N$ azimuthal angles
in the $N$ orthogonal spatial 2-planes, each with period $2\pi$. The remaining spatial dimensions
are parametrized by a radial coordinate $r$ and by $N +\epsilon = n = [D/2]$ ``direction cosines''
$\mu_i$ subject to the constraint $\sum_{i=1}^{N +\epsilon} \mu_i^2 = 1$, where $0\leq \mu_i \leq
1$ for $1\leq i\leq N$, and (for even $D$) $-1\leq \mu_{N+1}\leq 1$, $a_{N+1} = 0$.

The general nonextremal rotating, charged KK-AdS solutions can be elegantly written in a unified
form very similar to the Kerr-Schild metric ansatz \cite{KS65} by
\bea
ds^2 &=& H^{\frac{1}{D-2}}\Big(d\bar{s}^2 +\frac{2m}{UH}K^2\Big) \, , \label{KSm} \\
A &=& \frac{2ms}{UH}\, K \, , \qquad \Phi = \frac{-1}{D-2}\ln(H) \, , \label{KSgp}
\eea
where the anti-de Sitter metric $d\bar{s}^2$ and the timelike 1-form $K$ are given by
\bea
d\bar{s}^2 &=& -\big(1+g^2r^2\big)W\, d\bar{t}^2 +F\, dr^2 \nn \\
 && +\sum_{i=1}^{N +\epsilon}\frac{r^2+a_i^2}{\chi_i}\, d\mu_i^2
 +\sum_{i=1}^N\frac{r^2+a_i^2}{\chi_i}\mu_i^2\, d\bar{\phi}_i^2 \nn \\
&& -\frac{g^2}{\big(1+g^2r^2\big)W}\bigg( \sum_{i=1}^{N +\epsilon} \frac{r^2+a_i^2}{\chi_i}
 \mu_i\, d\mu_i \bigg)^2 \, , \\
K &=& c W\, d\bar{t} +\sqrt{f(r)}F\, dr
 -\sum_{i=1}^N\frac{a_i\sqrt{\Xi_i}}{\chi_i}\mu_i^2\, d\bar{\phi}_i \, , \quad
\eea
in which the functions ($U, W, F, H$) and $f(r)$ are defined to be
\bea
&& U = r^{\epsilon}\sum_{i=1}^{N +\epsilon}\frac{\mu_i^2}{r^2+a_i^2}\prod_{j=1}^N\big(r^2+a_j^2\big) \, ,
\qquad W = \sum_{i=1}^{N +\epsilon}\frac{\mu_i^2}{\chi_i} \, , \nn \\
&& F = \frac{r^2}{1+g^2r^2}\sum_{i=1}^{N +\epsilon}\frac{\mu_i^2}{r^2+a_i^2} \, , \qquad
 H = 1 +\frac{2ms^2}{U} \, , \nn \\
&&\qquad\qquad f(r) = c^2 -s^2\big(1+g^2r^2\big) \, , \nn \\
&& \qquad\quad \Xi_i = c^2 -s^2\chi_i \, , \qquad \chi_i = 1 -g^2a_i^2 \, . \nn
\eea
In the above and below, the shorthand notations $c = \cosh\delta$ and $s = \sinh\delta$ are used.

One may transform the above solutions to a frame in terms of generalized Boyer-Lindquist coordinates by
\be\begin{split}
d\bar{t} &= dt +\frac{2mc\sqrt{f(r)}\, dr}{(1+g^2r^2)[V(r) -2mf(r)]} \, , \\
d\bar{\phi}_i &= d\phi_i +\frac{2ma_i\sqrt{\Xi_i}\sqrt{f(r)}\, dr}{(r^2+a_i^2)[V(r) -2mf(r)]} \, ,
\end{split}\ee
where the function $V(r)$ is defined by
\bea
V(r) \equiv \frac{U}{F} = r^{\epsilon-2}\big(1+g^2r^2\big)\prod_{i=1}^N\big(r^2+a_i^2\big) \, . \nn
\eea
The general KK-AdS solutions then have the form
\bea
ds^2 &=& H^{\frac{1}{D-2}}\bigglb[ -\big(1+g^2r^2\big)W\, dt^2 +\frac{U\, dr^2}{V(r) -2mf(r)} \nn \\
&& +\sum_{i=1}^{N +\epsilon}\frac{r^2+a_i^2}{\chi_i}\, d\mu_i^2
 +\sum_{i=1}^N\frac{r^2+a_i^2}{\chi_i}\mu_i^2\, d\phi_i^2 \nn \\
&& -\frac{g^2}{\big(1+g^2r^2\big)W}\bigg( \sum_{i=1}^{N +\epsilon}\frac{r^2+a_i^2}{\chi_i}\mu_i\, d\mu_i \bigg)^2 \nn \\
&& +\frac{2m}{UH}\bigg( cW\, dt -\sum_{i=1}^N\frac{a_i\sqrt{\Xi_i}}{\chi_i}\mu_i^2\, d\phi_i \bigg)^2\biggrb] \, ,
 \label{KKAdSbh} \\
A &=& \frac{2ms}{UH}\bigg( cW\, dt -\sum_{i=1}^N\frac{a_i\sqrt{\Xi_i}}{\chi_i}\mu_i^2\, d\phi_i
\bigg) \, , \label{KKAdSg}
\eea
which is in a frame nonrotating at infinity. The gauge potential has been changed modulo a radial gauge
transformation.

In the uncharged case ($\delta = 0$), the above metric (\ref{KKAdSbh}) reduces to those found in
\cite{GLPP04,GLPP05}. On the other hand, if the cosmological constant is set to zero, it degenerates
to those derived in \cite{KMNV06}. In particular, the KK-AdS solutions in the $D = 4, 5, 7$ nonrotating
case correspond to the supergravity black hole solutions \cite{MC457} but with only one charge. I have
directly and explicitly checked that the general solutions (\ref{KKAdSbh}) and (\ref{KKAdSg}) obey the
field equations derived from the Lagrangian of the Einstein-Maxwell-dilaton system ($\mathcal{F} = dA$)
\begin{align}
\mathcal{L} &= \sqrt{-g}\Big\{ R -\frac{1}{4}(D-1)(D-2)(\p\Phi)^2
 -\frac{1}{4}e^{-(D-1)\Phi}\mathcal{F}^2 \nn \\
&\qquad +g^2(D-1)\big[(D-3)e^{\Phi} +e^{-(D-3)\Phi}\big]\Big\} \, ,
\end{align}
for the $D = 4, 5, 6, 7$ cases. Since the dimension of spacetime is not distinguished in any way in
the general expressions for the solutions, it is confident that the solutions are valid in all dimensions.
In particular, the $D = 4$ solution reproduces the one recently found by Chow \cite{Chow11}. In the
$D = 5$ case with only one rotation parameter and when one sets $\Xi_a = \Xi_b = 1$ here, the single-charged
solution then coincides with the special solution found in \cite{CCLP05} after setting $w = 1$ there.
However, in the case of rotating single-charged KK-AdS$_5$ solution with two unequal rotation parameters,
no simple relation has been found to make the above solution contact with the general solution declared
in \cite{CCLP07}.

\section{Thermodynamics}
The KK-AdS black holes have Killing horizons at $r = r_+$, the largest positive root of $V(r_+) = 2mf(r_+)$.
On the horizon, the Killing vector
\bea
l = \frac{\p}{\p t} +\sum_{i=1}^N\frac{\big(1+g^2r_+^2\big)a_i\sqrt{\Xi_i}}{\big(r_+^2+a_i^2\big)c}
 \, \frac{\p}{\p\phi_i} \nn
\eea
becomes null and obeys $l^{\mu}l_{\nu;\mu} = \kappa l_{\nu}$, where the surface gravity is given by
\bea
\kappa &=& \frac{r_+\big(1+g^2r_+^2\big)\sqrt{f(r_+)}}{c}\bigg[\sum_{i=1}^N\frac{1}{r_+^2+a_i^2} \nn \\
&& +\frac{\epsilon-2}{2r_+^2} +\frac{g^2c^2}{\big(1+g^2r_+^2\big)f(r_+)} \bigg] \, ,
\eea
hence the Hawking temperature $T = \kappa/(2\pi)$ is
\be
T = \frac{\sqrt{f(r_+)}\big[V'(r_+) -2mf'(r_+)\big]}{4\pi r_+^{\epsilon-2}c
 \prod_{i=1}^N\big(r_+^2+a_i^2\big)} \, .
\ee

The entropy of the outer horizon is easily evaluated as
\be
S = \frac{\mathcal{V}_{D-2}r_+^{\epsilon-1}c}{4\sqrt{f(r_+)}}
 \prod_{i=1}^N\frac{r_+^2+a_i^2}{\chi_i}
 = \frac{\mathcal{V}_{D-2}mr_+c\sqrt{f(r_+)}}{2\big(1+g^2r_+^2\big)\prod_{i=1}^N\chi_i} \, ,
\ee
where I denote the volume of the unit $(D-2)$-sphere as
\be
\mathcal{V}_{D-2} = \frac{2\pi^{(D-1)/2}}{\Gamma[(D-1)/2]} \, .
\ee

On the horizon, the angular velocities and the electrostatic potential $\Phi_+ = \ell^\mu A_\mu\big|_{r_+}$
are given by
\be
\Omega_i = \frac{\big(1+g^2r_+^2\big)a_i\sqrt{\Xi_i}}{\big(r_+^2+a_i^2\big)c} \, , \qquad
\Phi_+ = \frac{s}{c}\big(1+g^2r_+^2\big) \, .
\ee

I then adopt the procedure that was used in \cite{Koga05,CLP06a} to calculate the conserved charges
as follows,
\be\begin{split}
& M = \frac{\mathcal{V}_{D-2}m}{8\pi\prod_{j=1}^N\chi_j}\bigg[c^2\bigg(\sum_{i=1}^N\frac{2}{\chi_i}
 +\epsilon -2\bigg) +1 \bigg] \, , \\
& J_i = \frac{\mathcal{V}_{D-2}ma_ic\sqrt{\Xi_i}}{4\pi\chi_i\prod_{j=1}^N\chi_j} \, , \qquad
 Q = \frac{(D-3)\mathcal{V}_{D-2}mcs}{8\pi\prod_{j=1}^N\chi_j} \, ,
\end{split}\ee
and explicitly verify that they satisfy the differential and integral first laws of thermodynamics
\begin{subequations}\begin{align}
& dM = T\, dS +\sum_{i=1}^N\Omega_i\, dJ_i +\Phi_+\, dQ -P\, d\mathcal{V} \, , \\
& \frac{D-3}{D-2}(M -\Phi_+\, Q) = T\, S +\sum_{i=1}^N\Omega_i\, J_i -P\, \mathcal{V} \, ,
\end{align}\end{subequations}
where I have introduced the generalized pressure
\bea
P &=& \frac{g^{D-2}m}{4\pi(D-2)\prod_{j=1}^N\chi_j}\bigg[c^2\bigg(\sum_{i=1}^N\frac{1}{\chi_i}
 +\frac{D-3+\epsilon}{2} \nn \\
&&\quad -\frac{D-2}{1+g^2r_+^2}\bigg) -\frac{D-3}{2}s^2g^2r_+^2 \bigg] \, ,
\eea
which is conjugate to the volume $\mathcal{V} = \mathcal{V}_{D-2}g^{2-D}$ of the $(D-2)$-sphere with
the AdS radius $1/g$. The results presented above include those given in \cite{KMNV06} and \cite{GPP05}
as special cases when $g = 0$ and $\delta = 0$, respectively.

\section{Inclusion of the NUT charges}
To include the NUT charges, it is convenient to adopt the Jacobi-Carter coordinates used in \cite{CLP06b}.
In doing so, I find that the general nonextremal KK-AdS-NUT solutions can be cast into the following compact form:
\bea
ds^2 &=& H^{\frac{1}{D-2}}\bigg\{ -\prod_{\beta=1}^n \frac{1-g^2x_{\beta}^2}{\chi_{\beta}}\, dt^2
 +\sum_{\alpha=1}^n\frac{U_{\alpha}}{X_{\alpha}}\, dx_{\alpha}^2 \nn \\
&& +\sum_{i=1}^{n-\epsilon}\frac{\tilde{\mu}_i^2}{\chi_i}\, d\phi_i^2
 +\sum_{\alpha=1}^n\frac{2m_{\alpha}(-x_{\alpha})^{\epsilon}}{U_{\alpha}}K_{\alpha}^2 \nn \\
&& -\frac{s^2}{H}\bigg[\sum_{\alpha=1}^n\frac{2m_{\alpha}(-x_{\alpha})^{\epsilon}}{U_{\alpha}}
 K_{\alpha}\bigg]^2\bigg\} \, , \\
A &=& \frac{s}{H}\sum_{\alpha=1}^n\frac{2m_{\alpha}
 (-x_{\alpha})^{\epsilon}}{U_{\alpha}}K_{\alpha} \, , \\
\Phi &=& \frac{-1}{D-2}\ln(H) \, ,
\eea
with $n$ independent 1-forms
\be
K_{\alpha} = \frac{c}{1-g^2x_{\alpha}^2}\prod_{\beta=1}^n
 \frac{1-g^2x_{\beta}^2}{\chi_{\beta}}\, dt
 -\sum_{i=1}^{n-\epsilon}\frac{a_i\tilde{\mu}_i^2
 \sqrt{\Xi_i}}{\big(a_i^2-x_{\alpha}^2\big)\chi_i}\, d\phi_i \, .
\ee
In the above, I denote
\bea
&&\tilde{\mu}_i^2 = \frac{\prod_{\alpha=1}^n\big(a_i^2 -x_{\alpha}^2\big)}{{\prod'}_{k=1}^n
 \big(a_i^2 -a_k^2\big)} \, ,
\qquad U_{\alpha} = {{\prod}'}_{\beta=1}^n \big(x_{\beta}^2 -x_{\alpha}^2\big) \, , \nn \\
&& H = 1 +s^2\sum_{\alpha=1}^n\frac{2m_{\alpha}(-x_{\alpha})^{\epsilon}}{U_{\alpha}} \, , \quad
 f(x_{\alpha}) = 1 +g^2s^2x_{\alpha}^2 \, , \nn \\
&&\qquad X_{\alpha} = \bar{X}_{\alpha} +2m_{\alpha}(-x_{\alpha})^{\epsilon}f(x_{\alpha}) \, , \nn \\
&&\qquad \bar{X}_{\alpha} = \frac{1-g^2x_{\alpha}^2}{x_{\alpha}^2}
 \prod_{i=1}^n\big(a_i^2 -x_{\alpha}^2\big) \, , \nn
\eea
where the prime on the product symbol in the definition of $\tilde{\mu}_i^2$ or $U_{\alpha}$
indicates that the vanishing factor (i.e. when $k = i$ or $\alpha = \beta$) is to be omitted.
In odd dimensions, one has $m_n = m$; and for even dimensions, $m_n = im$ and $a_n = 0$. In
all dimensions, one sets $x_n = ir$.

One may transform the above KK-AdS-NUT metrics via the following coordinate transformations,
\be\begin{split}
d\bar{t} &= dt +c\sum_{\alpha=1}^n\frac{2im_{\alpha}(-x_{\alpha})^{\epsilon}
 \sqrt{f(x_{\alpha})}}{\big(1-g^2x_{\alpha}^2\big)X_{\alpha}}\, dx_{\alpha} \, , \\
d\bar{\phi}_j &= d\phi_j +a_j\sqrt{\Xi_j}\sum_{\alpha=1}^n \frac{2im_{\alpha}
 (-x_{\alpha})^{\epsilon}\sqrt{f(x_{\alpha})}}{\big(a_j^2
 -x_{\alpha}^2\big)X_{\alpha}}\, dx_{\alpha} \, ,
\end{split}\ee
into a form like the multi-Kerr-Schild ansatz \cite{CL07}. The expected line element and the
gauge potential can be rewritten as follows
\bea
ds^2 &=& H^{\frac{1}{D-2}}\bigg\{d\bar{s}^2 +\sum_{\alpha=1}^n\frac{2m_{\alpha}
 (-x_{\alpha})^{\epsilon}}{U_{\alpha}}\bar{K}_{\alpha}^2 \nn \\
&& -\frac{s^2}{H}\bigg[\sum_{\alpha=1}^n \frac{2m_{\alpha}
 (-x_{\alpha})^{\epsilon}}{U_{\alpha}}\bar{K}_{\alpha}\bigg]^2\bigg\} \, , \\
A &=& \frac{s}{H}\sum_{\alpha=1}^n\frac{2m_{\alpha}
 (-x_{\alpha})^{\epsilon}}{U_{\alpha}}\bar{K}_{\alpha} \, ,
\eea
where the NUT-AdS metric $d\bar{s}^2$ and the timelike 1-forms $\bar{K}_{\alpha}$ read
\bea
d\bar{s}^2 &=& -\prod_{\beta=1}^n \frac{1-g^2 x_{\beta}^2}{\chi_{\beta}}\, d\bar{t}^2
 +\sum_{\alpha=1}^n\frac{U_{\alpha}}{\bar{X}_{\alpha}}\, dx_{\alpha}^2
 +\sum_{i=1}^{n-\epsilon}\frac{\tilde{\mu}_i^2}{\chi_i}\, d\bar{\phi}_i^2 \, , \nn \\
&& \\
\bar{K}_{\alpha} &=& \frac{c}{1-g^2x_{\alpha}^2}
 \prod_{\beta=1}^n\frac{1-g^2x_{\beta}^2}{\chi_{\beta}}\, d\bar{t}
 +i\sqrt{f(x_{\alpha})}\frac{U_{\alpha}}{\bar{X}_{\alpha}}\, dx_{\alpha} \nn \\
&& -\sum_{i=1}^{n-\epsilon}\frac{a_i\sqrt{\Xi_i}}{\big(a_i^2
 -x_{\alpha}^2\big)\chi_i}\tilde{\mu}_i^2\, d\bar{\phi}_i \, .
\eea
Reinterpreted in $D+1$ dimensions, the general KK-AdS-NUT solutions can be obtained via
the standard dimension reduction along the $z$-direction from the $(D+1)$-dimensional
line element
\be
d\hat{s}^2 = dz^2 +d\bar{s}^2 +\sum_{\alpha=1}^n\frac{2m_{\alpha}
 (-x_{\alpha})^{\epsilon}}{U_{\alpha}}\big(\bar{K}_{\alpha} +s\, dz\big)^2 \, .
\ee

\section{Metric structure and MOND}
Expressed in the language of tensors, the general nonextremal KK-AdS solutions (\ref{KSm}) and (\ref{KSgp}) have
a beautiful structure as follows
\be\begin{split}
g_{\mu\nu} &= H^{\frac{1}{D-2}}\Big(\bar{g}_{\mu\nu} +\frac{2m}{UH}K_{\mu}K_{\nu}\Big) \, , \\
g^{\mu\nu} &= H^{\frac{-1}{D-2}}\Big(\bar{g}^{\mu\nu} -\frac{2m}{U}K^{\mu}K^{\nu}\Big) \, , \\
A_{\mu} &= \frac{2ms}{UH} K_{\mu} \, , \label{KKAdSms}
\end{split}\ee
with  the vector $K^{\mu} = \bar{g}^{\mu\nu}K_{\nu}$ being raised by the background metric tensors
$\bar{g}^{\mu\nu}$. Moreover, the vector $K_{\mu}$ is a timelike geodesic congruence, satisfying
\be
\bar{g}^{\mu\nu}K_{\mu}K_{\nu} = -s^2 \, , \quad
K^{\mu}\bar{\nabla}_{\mu}K_{\nu} = K^{\mu}\bar{\nabla}_{\nu}K_{\mu} = 0 \, .
\ee

In the presence of ($n-1$) NUT charges and the mass parameter, with the help of the relations:
$\bar{g}_{\mu\nu}
\bar{K}_{\alpha}^{\mu}\bar{K}_{\alpha}^{\nu} = \bar{g}_{\mu\nu}\bar{K}_{\alpha}^{\mu}
\bar{K}_{\beta}^{\nu} = -s^2$, the inverse metric is found to be
\be
 g^{\mu\nu} = H^{\frac{-1}{D-2}}\bigg[\bar{g}^{\mu\nu} -\sum_{\alpha=1}^n\frac{2m_{\alpha}
 (-x_{\alpha})^{\epsilon}}{U_{\alpha}}\bar{K}_{\alpha}^{\mu}\bar{K}_{\alpha}^{\nu}\bigg] \, .
\ee

The metric structure (\ref{KKAdSms}) naturally reduces to the notable Kerr-Schild ansatz \cite{KS65}
in the uncharged case, therefore it is likely the unique suitable generalization to that of the Kaluza-Klein
gauged and ungauged supergravity theory in the case with only one electric charge. It has already been
checked that with some further modifications, the ansatz (\ref{KKAdSms}) can be extended to large classes
of already-known black hole solutions with multiple pure electric charges, both in the cases of rotating
charged black holes in ungauged supergravity and in the cases of nonrotating AdS black holes in gauged
supergravity. Guided by the generalized ansatz, in principle, one is able to construct the expected new
exact gauged solutions. Therefore, it would be highly expected that the generalized ansatz (\ref{KKAdSms})
can open a new way towards constructing the most general rotating charged AdS black hole solutions with
multiple pure electric charges in gauged supergravity theory.

Alternatively, one may use the dilaton scalar to reexpress the metric tensors and the gauge potential as
\be\begin{split}
g_{\mu\nu} &= e^{-\Phi}\bar{g}_{\mu\nu} +\big[e^{-\Phi} -e^{(D-3)\Phi}\big]s^{-2}K_{\mu}K_{\nu} \, , \\
g^{\mu\nu} &= e^{\Phi}\bar{g}^{\mu\nu} +\big[e^{\Phi} -e^{-(D-3)\Phi}\big]s^{-2}K^{\mu}K^{\nu} \, , \\
A_{\mu} &= \big[1 -e^{(D-2)\Phi}\big]s^{-1}K_{\mu} \, .
\end{split}\ee
It is apparent that the metric of the four-dimensional KK-AdS black holes resembles those proposed by Yilmaz
\cite{WCdS11} and by Bekenstein \cite{JDB04}. Given the same matter contents of these two tensor-vector-scalar
(TeVeS) theories, it deserves a lot of deeper investigations of the relation between them and the astrophysical
implications of the four-dimensional Kaluza-Klein-AdS theory as another kind of modified Newtonian dynamics
(MOND). For instance, it is an interesting question as to whether the experimental test of effects of the
four-dimensional Kaluza-Klein black hole on our solar system can explore the existence of extra spatial
dimensions or put some constrains on the size of extra fifth dimension.

\section{Conclusions}
In this paper, I have found the general nonextremal rotating, charged KK-AdS black holes with arbitrary
angular momenta in all higher dimensions, and extended them to include the ($n-1$) NUT charges. The
conserved charges are given explicitly and shown to obey the differential and integral first laws of
black hole thermodynamics. I then have exploited that the general nonextremal KK-AdS solutions have a
beautiful structure similar to the Kerr-Schild ansatz, which highlights its promising application to
include multiple electric charges into solutions yet to be discovered in gauged supergravity. In addition,
it is also observed that the generalized ansatz in the $D = 4$ case can be expressed as a form like the
one previously suggested by Yilmaz and later by Bekenstein in his TeVeS theory.

\begin{acknowledgments}
S.-Q. Wu is supported by the NSFC under Grant Nos. 10975058 and 10675051. The computation within this
work has been done by using the GRTensor-II program based on Maple 7.
\end{acknowledgments}

\medskip
\textit{Notes added}.
While this paper was being under review (and finally rejected by PRL), a generalized form of the ansatz
proposed in this work was used to successfully construct a new exact rotating charged solution with two
unequal rotation parameters and with two different electric charges, but the third charge being set to
zero, in five-dimensional $U(1)^3$ gauged supergravity. Along the same line, the present author now succeeds
in constructing the most general charged rotating AdS$_5$ solution with three unequal charges and with two
independent rotation parameters, which is the most interesting solution previously unknown in $D = 5$
$U(1)^3$ gauged supergravity. As such, it is believed that the ansatz (\ref{KKAdSms}) proposed in this
paper and its generalized form would bring about an important breakthrough in the method of constructing
new exact gauged supergravity solutions.

\end{document}